\documentclass[prd,a4paper,showpacs,showkeys,byrevtex]{revtex4}
\begin{document}

\title{Electromagnetic splitting for mesons and baryons \\
using dressed constituent quarks}
\author{Bernard \surname{Silvestre-Brac}}
\email[E-mail: ]{silvestre@lpsc.in2p3.fr}
\affiliation{Laboratoire de Physique Subatomique et de Cosmologie,
53, Av. des Martyrs, F-38026 Grenoble-Cedex, France}
\author{Fabian \surname{Brau}}
\thanks{FNRS Postdoctoral Researcher}
\email[E-mail: ]{fabian.brau@umh.ac.be}
\author{Claude \surname{Semay}}
\thanks{FNRS Research Associate}
\email[E-mail: ]{claude.semay@umh.ac.be}
\affiliation{Universit\'{e} de Mons-Hainaut, Place du Parc, 20,
B-7000 Mons, Belgium}
\date{\today}

\begin{abstract}
Electromagnetic splittings for mesons and baryons are calculated in a
formalism where the constituent quarks are considered as dressed
quasiparticles. The electromagnetic interaction, which contains coulomb,
contact, and hyperfine terms, is folded with the quark electrical
density. Two different types of strong potentials are considered.
Numerical treatment is done very carefully and several approximations
are discussed in detail. Our model contains only one free parameter and
the agreement with experimental data is reasonable although it seems
very difficult to obtain a perfect description in any case.
\end{abstract}



\maketitle

\section{Introduction}

Quantum Chromodynamics (QCD) is believed to be the good theory of
strong interaction. It has met with numerous successes in many domains.
However, in the low energy regime, it is extremely difficult to handle
because of its non perturbative character. Lattice calculations become
more and more reliable but still remain very cumbersome, time consuming
and not always transparent for the underlying physics. This explains
why, in the meson and baryon sectors, a number of alternative simpler
models were introduced. Among them, the non relativistic quark model
(NRQM) is very appealing because of its high simplicity, its ability to
treat properly the center of mass motion, and the large number of
observables that can be described within its framework.

In NRQM formalism, the dynamical equation is the usual Schr\"{o}dinger
equation including a non relativistic kinetic energy term plus a
potential term \cite{luch91}. There exist a lot of different numerical
algorithms to solve the two-body and three-body problems with a good
accuracy (see for instance Refs.~\cite{baye86,baye97,suzu98}).

Nowadays, it becomes more and more frequent to use a relativistic
expression for the kinetic energy operator. The resulting dynamical
equation is known as a spinless Salpeter equation. It has several
advantages as compared to the Schr\"{o}dinger equation
\cite{sema92,fulc94} and the corresponding numerical algorithms are now
well under control (see for instance
Refs.~\cite{suzu98,fulc94,brau98b,sema01,silv01}). This kind of models
and NRQM
both use potentials, and are called potential models.

In potential models (and indeed in many other QCD inspired models), the
quark
degrees of freedom are no longer the bare quarks of the QCD lagrangian,
but are quasiparticles dressed by a gluon cloud and quark-antiquark
virtual pairs. They are called constituent quarks and the most visible
modification is the necessity to use in potential models a quark mass
substantially
larger than the bare mass. In principle the bare quark-quark potential
should also be modified and folded with the quark color density to give
the final potential to be used in a Schr\"{o}dinger or in a spinless
Salpeter equation. In recent quark-quark potentials appearing on the
market, this effect is taken into account \cite{sema97,sema99,brau98,
sema01b}. Actually, this effect has been already considered since a
rather long time (see i.e. Ref.~\cite{godf85,caps86}). The resulting
spectra of these models are in correct agreement with experimental data.
However, it seems very difficult to get, in a unified scheme (same form
of the potential and same set
of parameters), a good description of both meson and baryon properties
\cite{blas90,isgu00}. For example in the well known works of Isgur
\cite{godf85,caps86}, the string tension is different in the mesonic and
in the baryonic sector. Recently, progresses in obtaining a unified
description have been achieved \cite{brau02}.

The spectra are only a part of interesting observables, and the validity
of a model should be tested on other observables, especially if they are
very sensitive to the form of the wave function. Electromagnetic
properties are best suited for such a study, because the basic QED
formalism is very well known and precise, and thus the possible
uncertainties coming from mechanisms or wave function (itself depending
on the much less known strong interactions) are more conveniently
identified.

In this paper, we focus our study on the electromagnetic splitting
between charged hadrons, both in the meson and baryon sectors, within
the framework of potential models. Since the earliest works on charmed
mesons
\cite{john78}, and baryons \cite{wrig78}, a number
of similar studies were performed in the past
\cite{luch91,brau98,godf86,caps87,geno98}.
Essentially three different sources for the splitting were identified: a
small mass difference between up and down quarks, the coulomb
interaction between charged quarks, and their dipole-dipole interaction
\cite{geno98}.
All of them seem to have an important effect and the final result is a
very subtle interplay among them. This explains why a very proper and
precise treatment must be invoked, and also why this observable is very
interesting.
The current mass ratio for the $u$ and $d$ quarks is probably comprised
between 0.2 and 0.8 \cite{pdg}. However their absolute values are
presumably weak (a few MeV) with respect to the QCD scale parameter
$\Lambda$. Thus the spontaneous symmetry breaking induces large
constituent masses for the $u$ and $d$ quarks; the corresponding values
are very close (making the SU(2) isospin symmetry rather good) but
nevertheless different. This small difference is a first source for the
isospin splitting.
Moreover, quarks being charged particles, the coulomb interaction
is obviously present (very often, it has been treated as a
perturbation). The dipole-dipole interaction (or hyperfine interaction)
is a consequence of relativistic corrections to the coulomb potential.

The same problem was also undertaken using formalisms relying more
basically on fundamental QCD. In Ref. \cite{luty95}, an heavy quark
effective theory (first order in $1/m_Q$) is adopted to study the
splitting in heavy mesons, using dispersion relations. In
Ref. \cite{gao97} a chiral field theory is employed to study several
splittings in ordinary and strange sectors. In Ref. \cite{cola98}, the
authors are interested in the charmed and bottom meson sectors with a
formalism based on Cottingham formula. In Ref. \cite{delb99} a tadpole
term is introduced to deal with the $u-d$ mass
difference in some hadrons. But in most of these studies, the authors
limit themselves to very restricted samples, either in meson or in
baryon sector. We think that so few states to test the validity of
a model can be questionable.

In our paper we want to deal with \emph{all the known splittings} both
in mesons and in baryons in a consistent approach, and to push the
potential model study further in several domains.
First we want to perform a precise and complete treatment, avoiding
perturbative expressions as much as possible. Second, we introduce the
``contact term'', which arises on equal footing as the dipole-dipole
relativistic correction, but which is neglected by most authors
\cite{luch91,geno98}.

Finally, our most important improvement to our
point of view is the use of a dressed electromagnetic interaction
between quarks. Since the constituent quarks are quasiparticles, the
electromagnetic interaction should also be modified as compared to the
bare one, in a very similar way to the quark-quark strong potential.
However the electromagnetic lagrangian is different from the QCD
lagrangian and the electromagnetic density for the quark, playing a role
in the splitting, has no reason to be identical to the color density
occurring in the quark-quark strong potential.
Such an approach has already been proposed in Refs.~\cite{godf86,caps87}
but with a different form for the electromagnetic density. Moreover, in
these works, different sets of parameters have been used for mesonic and
baryonic sectors separately.

In order to see the sensitivity of the results on the treatment of the
strong interactions for the quark dynamics,
we investigate the splitting produced with two types of wave function,
one resulting from a phenomenological non relativistic hamiltonian (AL1)
\cite{silv93,Sema94} and another with a semi-relativistic hamiltonian
(called here BSS) \cite{brau02}. Moreover, since our aim is to consider
mesons and baryons on equal footing, it is important to consider
interquark potentials that lead to a correct description of both
sectors. This is rather difficult to encounter. Both hamiltonians
considered here are suited for that. The first potential (AL1) relies on
the so-called funnel or Cornell potential \cite{eich75,bhad81}. It is
completely phenomenological and one can consider that the dressing of
the quarks is included and simulated in the value of the various
parameters. The second one (BSS) starts with more fundamental QCD
grounds and is based on instanton induced effects \cite{blas90,shur89};
the dressing is explicitly taken into account but there remain
nevertheless some free parameters that are adjusted on the spectra.

The paper is organized as follows. The next section presents in more
details the strong potentials and the way to solve the two-body and
three-body problems. The third section deals with the electromagnetic
interaction responsible of the splitting. The results of our
calculations are presented and compared to data in the fourth section.
Conclusions are drawn in the last section.

\section{Potentials and wave function}

\subsection{Strong potentials}

As stated in the introduction, we use in this paper two kinds of
interquark potentials: one that must be used in a Schr\"{o}dinger
equation (AL1) and the other in a spinless Salpeter equation (BSS). They
both depend on the relative distance $r$ between the interacting quarks
and are able to describe in a satisfactory way both the meson and the
baryon spectra. However the BSS potential is suited,
because of its underlying QCD basis, only for the light quark sectors
($u$, $d$, $s$ quarks). They differ by the type of
kinematics and by the manner to deal with spin splitting; although they
give spectra of similar good quality, the corresponding wave functions
can differ appreciably. Their derivation and their
parameters have been reported elsewhere, and here we just want to point
out the essential features and stress their differences more explicitly.
In both models, the $u$ and $d$ quarks are assumed to have the same
mass. In the following, they will be noted by the symbol $n$ (normal or
non strange).

\subsubsection{AL1 Potential}

The AL1 potential \cite{Sema94}, developed for a non relativistic
kinematics, contains the minimum ingredients
necessary to get an overall reasonable description of hadronic
resonances, a central part $V_{C}$ and a hyperfine term $V_{H}$
\begin{equation}
V_{ij}(r)=-\frac{3}{16}{\bf \lambda}_{i}\cdot {\bf \lambda}_{j}
\left[V_{C}^{(ij)}(r)+V_{H}^{(ij)}(r)\right].
\label{pot}
\end{equation}
The central part is merely the Cornell potential composed of a short
range coulombic part, simulating the one-gluon exchange mechanism, and a
long range linear term,
responsible for the confinement (an additional
constant is very important to get the good absolute values)
\begin{equation}
V_{C}^{(ij)}(r)=-\frac{\kappa}{r}+ar+C.
\end{equation}
The color dependence through the Gell-Mann matrices $\lambda$ in
relation~(\ref {pot}) comes from one gluon exchange. There is no reason,
except simplicity, that such a structure is kept for the confining and
constant parts. Nevertheless, this ansatz works well for both meson and
baryon sectors.

The hyperfine term has a short range behavior and is chosen as a
gaussian function
\begin{equation}
V_{H}^{(ij)}(r)=\frac{8\pi }{3m_{i}m_{j}}\kappa ^{\prime }\frac{\exp
(-r^{2}/\sigma_{ij}^{2})}{\pi ^{3/2}\,\sigma_{ij}^{3}}{\bf s}_{i} \cdot
{\bf s}_{j}.
\label{hypal1}
\end{equation}
The interesting property, as compared to Bhaduri's \cite{bhad81} or
Cornell's \cite{eich75} potentials, is that the range $\sigma_{ij}$ of
that force does depend on the flavor
\begin{equation}
\sigma_{ij}=A\left( \frac{2m_{i}m_{j}}{m_{i}+m_{j}}\right)^{-B}.
\end{equation}
A kind of dressing is realized with the gaussian function, since the
theory predicts a delta contribution for the hyperfine potential.
The various parameters, including constituent masses, have been
determined on the spectra by a best fit procedure. Although very simple,
this potential does already a good job in hadronic spectroscopy.

The AL1 potential depends on quark masses in the strong hyperfine term.
So a variation of the quark masses, as necessary for treating the
electromagnetic splitting, has a strong effect on the meson masses. We
have checked that this
may induce wrong results in the case of a perturbative calculation.

\subsubsection{BSS Potential}
\label{sec:bss}

The BSS potential \cite{brau02}, developed for a semi relativistic
kinematics, contains the Cornell potential and an instanton induced
interaction. The Cornell potential has the same form as in the AL1
potential, but the constant interaction is different for meson and
baryon sectors. Contrary to the AL1 potential, the instanton formalism
is based on a SU(3) flavor symmetry and can hardly be generalized to the
heavy quark sector.

The instanton induced interaction provides a suitable formalism to
reproduce well the spectrum of the pseudoscalar mesons (and to explain
the masses  of $\eta$ and $\eta'$ mesons). The interaction between one
quark and one antiquark in a meson is vanishing for $L\neq 0$ or
$S \neq 0$ states. For $L=S=0$, its form depends on the isospin of the
$q\bar q$ pair
\begin{itemize}
\item For $I=1$:
\begin{equation}
\label{ins2}
V_{I}(r)=-8\, g\, \delta(\vec{r}\,);
\end{equation}
\item For $I=1/2$:
\begin{equation}
\label{ins3}
V_{I}(r)=-8\, g'\, \delta(\vec{r}\,);
\end{equation}
\item $I=0$:
\begin{equation}
\label{ins4}
V_{I}(r)=8
\left(
\begin{array}{cc}
g & \sqrt{2}g' \\
\sqrt{2}g' & 0
\end{array}
\right)\, \delta(\vec{r}\,),
\end{equation}
in the flavor space $(1/\sqrt{2}(|u\bar{u} \rangle+|d\bar{d}
\rangle),|s\bar{s} \rangle)$.
\end{itemize}
The parameters $g$ and $g'$ are two
dimensioned coupling constants. Between two quarks in a baryon, this
interaction is written \cite{blas90,munz94}
\begin{equation}
\label{vinst}
V_{I}(r)=-4 \left(g P^{[nn]} + g' P^{[ns]} \right) P^{S=0}
\delta(\vec r\,),
\end{equation}
where $P^{S=0}$ is the projector on spin 0, and $P^{[qq']}$ is the
projector on antisymmetrical flavor state $qq'$.
We have shown in Ref.~\cite{brau02} that this model is able to
describe correctly meson and baryon spectra in a consistent way
if the following interaction is added for baryons only
\begin{equation}
\label{vinstnew}
V_{I}^{\rm baryon}(r)=C_{I} \left(P^{[nn]} + P^{[ns]} \right)
P^{S=0} P^{L=0},
\end{equation}
where $C_{I}$ is a new constant.

The quark masses used in this model are the constituent masses and not
the current ones. It is then natural to suppose that a quark is not a
pure pointlike particle, but an effective degree of freedom which is
dressed by the gluon and quark-antiquark pair clouds. The form that we
retain for the color charge density of a quark is a Gaussian function
\begin{equation}
\label{rhoi}
\rho({\bf r}) = \frac{1}{(\Gamma\sqrt{\pi})^{3/2}} \exp(- r^2 /
\Gamma^2).
\end{equation}
It is generally assumed that the quark size $\Gamma$ depends on the
flavor. So, we consider two size parameters $\Gamma_n$ and $\Gamma_s$
for $n$ and $s$ quarks respectively.
More details are given in Refs.~\cite{brau98,sema01b,brau02}.

The instanton interaction acts differently on the symmetric $(ud+du)$
and antisymmetric $(ud-du)$ flavor states; in a framework where
$m_{u}=m_{d}$, the isospin formalism can be introduced to classify the
basis states and this property can be taken into account without
problem. As soon as $m_{u}\neq m_{d}$, the up
and down quark must be considered as different and the instanton becomes
very difficult to handle. Thus, for baryons, the three-body
treatment with BSS is done perturbatively. In this case, the potential
has no mass dependence and the problem mentioned above for the AL1
potential does not appear.

\subsection{Numerical Techniques}

There exist many numerical algorithms to compute the radial function of
a meson. In this paper, we use the method based on Lagrange mesh, which
is very simple, very precise and very fast, and for which a recent work
has shown that relativistic kinematics can be handled without any
problem \cite{sema01}. Thus the same method can be applied for both AL1
and BSS potentials. To get a very high relative precision, around
$10^{-10}$, a typical number of basis states is $N=60$. Technical
details can be found in Ref.~\cite{sema01}.

Our method to solve the three-body problem is based on an expansion of
the space wave function in terms of harmonic oscillator functions with
\emph{two different sizes}. This method, which is an old one
\cite{nunb77}, was given up during a long time and renewed recently in
Ref.~\cite{silv01} (many authors use a numerical treatment based on
harmonic oscillators but with a \emph{single} variational parameter, for
instance in Ref.~\cite{caps86}). In this paper, it was shown that the
precision achieved was similar to the stochastic variational method
\cite{suzu98}. This last method was also shown to be competitive with
more conventional methods, such as Faddeev formalism \cite{roux95}. The
number of quanta is the relevant quantity for convergence; a very good
relative precision of the order of $10^{-5}$, necessary in our study, is
achieved if we include all basis states up to 20 quanta. More details on
the method can be found in
Ref.~\cite{silv01}.

\section{Electromagnetic interaction}

\subsection{Bare potential}

The electromagnetic interaction between two pointlike particles $i$ and
$j$ of charges $Q_{i}$, $Q_{j}$ and masses $m_{i}$, $m_{j}$ is very well
known. In addition to the usual coulomb potential $U_{\rm coul}$,
relativistic corrections (at lowest order) give rise to contact,
hyperfine, tensor, symmetric and antisymmetric spin-orbit, and Darwin
terms. Tensor, spin-orbit, and Darwin terms are complicated and
presumably their effects are weak, so that people neglect them. The
hyperfine interaction $U_{\rm hyp}$ does play an important role and is
included in all serious calculations. The contact term $U_{\rm cont}$ is
usually discarded with no other justification than simplicity. In this
study we keep it, so that our total bare electromagnetic potential
writes
\begin{equation}
U_{ij}^{(b)}({\bf r})=(U_{\rm coul})_{ij}^{(b)}({\bf r}%
)+(U_{\rm cont})_{ij}^{(b)}({\bf r})+(U_{\rm hyp})_{ij}^{(b)}({\bf r})
\label{potemb},
\end{equation}
with
\begin{eqnarray}
(U_{\rm coul})_{ij}^{(b)}({\bf r})&=& \phantom{+} Q_{i}Q_{j}
\frac{\alpha }{r},
\label{coulb} \\
(U_{\rm cont})_{ij}^{(b)}({\bf r})&=&-\frac{\pi }{2}Q_{i}Q_{j}
\left(\frac{1}{%
m_{i}^{2}}+\frac{1}{m_{j}^{2}}\right) \alpha \,\delta ({\bf r}),
\label{contb} \\
(U_{\rm hyp})_{ij}^{(b)}({\bf r})&=&-\frac{8\pi Q_{i}Q_{j}}{3m_{i}m_{j}}
\alpha \,\delta ({\bf r}){\bf s}_{i} \cdot {\bf s}_{j},
\label{hypb}
\end{eqnarray}
where $\alpha $ is the fine structure constant.
Let us stress a point that deserves further discussion. One can
wonder why such a contact term is not kept as well in the strong
potential AL1. This term comes from one gluon exchange and is
proportional to
$\alpha_s$; but this parameter is essentially phenomenological and the
Cornell form of the central potential is a kind of averaging of all
unknown effects coming from QCD, including the contact term. Our
philosophy is to maintain as far possible the form of the strong
potential, that was shown to do a good job in hadronic physics. On the
contrary, the situation is completely different in the electromagnetic
hamiltonian (\ref{potemb}) because the corresponding expression is the
exact one and does not suffer from ambiguities or phenomenological
uncertainties. This is why we decided to include the contact term, which
is usually neglected, in order to grasp quantitatively its effect and to
justify a posteriori the validity (or not) of neglecting it.

A perturbative treatment is compulsory for such a bare electromagnetic
potential, owing to the presence of the delta function. It was shown
that, in some cases, this a very bad approximation \cite{brau98}.

\subsection{Electromagnetic quark density}

One way to introduce the very complicated mechanisms that transform the
bare quarks (pointlike) to constituent quarks is to assume the existence
of a phenomenological density distribution for a quark. The
electromagnetic  density should in principle depend on the internal
structure of the system and contain a lot of unknown processes
(including probably interferences between QCD and QED) leading to
isospin symmetry breaking.

This means that
a constituent quark located at ${\bf r}$ is generated by a bare quark
located in ${\bf r}^{\prime } $ with a certain probability distribution
(or density) $\rho ({\bf r-r} ^{\prime })$. To have an appealing
physical meaning this density must be a peaked function around zero with
a certain size parameter, that depends in principle on the quark
flavor. Moreover we require the natural property that, for a vanishing
size, the constituent quark becomes pointlike and is identified to the
bare quark. Mathematically, this means that the limit of the density
$\rho ({\bf u})$ for a vanishing size is the delta function
$\delta ({\bf u)}$. Another natural property is that the density is
isotropic. The last required property is that the integral of
the density over the whole space is unity.

The most popular densities are of lorentzian, gaussian or Yukawa type.
For instance, the authors of Refs.~\cite{godf86,caps87} adopted a
gaussian form.
In this study we choose an electromagnetic Yukawa density. There is a
precise reason for that: this density is the leading ingredient of the
meson charge form factor. It is an experimental fact that
the data accommodate rather nicely to a Yukawa density, giving a form
factor with an asymptotic behavior $Q^{-2}$ \cite{brau01,silv02},
instead of a gaussian density which leads to gaussian asymptotic
behavior.
Keeping in mind the previous remarks, the adopted density for quark of
flavor $i$ is
\begin{equation}
\rho _{i}({\bf u})=\frac{1}{4\pi \gamma _{i}^{2}}
\frac{e^{-u/\gamma_{i}}}{u},
\label{dens}
\end{equation}
where $\gamma _{i}$ is the electromagnetic size parameter.

The dressed potential $U$ is obtained from the bare potential $U^{(b)}$
by a double convolution over the densities of each interacting quark
\begin{equation}
U_{ij}({\bf r})=\int d{\bf u}\,d{\bf v}\,\rho _{i}({\bf u}
)\,\rho _{j}({\bf v})\,U_{ij}^{(b)}({\bf r}+{\bf v}-{\bf u}).
\label{convd}
\end{equation}
With a trivial change of variable, it is quite easy to transform this
double folding into a single one
\begin{equation}
U_{ij}({\bf r})=\int d{\bf r}^{\prime }\;U_{ij}^{(b)}({\bf r}
^{\prime })\,\rho _{ij}({\bf r}-{\bf r}^{\prime }),
\label{convs}
\end{equation}
with the definition of the new density
\begin{equation}
\rho _{ij}({\bf u})=\int d{\bf v}\,\rho _{i}({\bf v})\,\rho _{j}(%
{\bf u}-{\bf v}).
\label{densij}
\end{equation}

Applying Eq.~(\ref{densij}) to the Yukawa density (\ref{dens}), one gets
\begin{equation}
\rho _{ij}({\bf u})=\frac{1}{8\pi \gamma _{i}^{3}}e^{-u/\gamma _{i}}
\quad {\rm if} \quad \gamma _{i}=\gamma _{j}
\label{dyuki}
\end{equation}
and
\begin{eqnarray}
\rho _{ij}({\bf u})&=&\frac{1}{4\pi (\gamma _{i}^{2}-\gamma _{j}^{2})}
\left( \frac{e^{-u/\gamma _{i}}}{u}-\frac{e^{-u/\gamma _{j}}}{u}\right)
\\
 &=&\frac{1}{(\gamma _{i}^{2}-\gamma _{j}^{2})}\left( \gamma _{i}^{2}
 \rho _{i}
({\bf u})-\gamma _{j}^{2}\rho _{j}({\bf u})\right) \quad {\rm if} \quad
\gamma _{i}\neq \gamma _{j}.
\label{dyukij}
\end{eqnarray}

\subsection{Dressed potential}

Starting from Eq.~(\ref{potemb}) and using Eqs.~(\ref{convs}) and
(\ref{dyuki}), it is easy to obtain analytically the dressed potential
\begin{equation}
U_{ij}(r)=(U_{\rm coul})_{ij}(r)+(U_{\rm cont})_{ij}(r) +
(U_{\rm hyp})_{ij}(r),
\label{potemd}
\end{equation}
in the case of two interacting constituent quarks with the same size
$\gamma_{i}=\gamma _{j}=\gamma $. It writes
\begin{eqnarray}
(U_{\rm coul})_{ij}(r)&=&\phantom{+}\alpha Q_{i}Q_{j}\left[ \frac{1}{r}
\left
(1-e^{-r/\gamma}\right) -\frac{e^{-r/\gamma }}{2\gamma }\right],
\label{couldi} \\
(U_{\rm cont})_{ij}(r)&=&-\frac{\alpha Q_{i}Q_{j}}{16\gamma ^{3}}\left(
\frac{1}{m_{i}^{2}}+\frac{1}{m_{j}^{2}}\right) \,e^{-r/\gamma},
\label{contdi} \\
(U_{\rm hyp})_{ij}(r)&=&-
\frac{\alpha Q_{i}Q_{j}}{3\gamma ^{3}m_{i}m_{j}}
\,e^{-r/\gamma }{\bf s}_{i}\cdot {\bf s}_{j}.
\label{hypdi}
\end{eqnarray}
Doing the same thing with Eq.~(\ref{dyukij}), one obtains the
electromagnetic dressed potential in the case of different interacting
quarks
\begin{eqnarray}
(U_{\rm coul})_{ij}(r)&=&\phantom{+}\alpha Q_{i}Q_{j}\left( \frac{1}{r}-
\frac{\gamma _{i}^{2}}
{\gamma _{i}^{2}-\gamma _{j}^{2}}\frac{e^{-r/\gamma _{i}}}{r}+
\frac{\gamma_{j}^{2}}{\gamma _{i}^{2}-\gamma _{j}^{2}}
\frac{e^{-r/\gamma _{j}}}{r}\right),
\label{couldd} \\
(U_{\rm cont})_{ij}(r)&=&-\frac{\alpha Q_{i}Q_{j}}{8(\gamma _{i}^{2}-
\gamma_{j}^{2})}\left( \frac{1}{m_{i}^{2}}+\frac{1}{m_{j}^{2}}\right)
\frac{e^{-r/\gamma _{i}}-e^{-r/\gamma _{j}}}{r},
\label{contdd} \\
(U_{\rm hyp})_{ij}(r)&=&-\frac{2\alpha Q_{i}Q_{j}}{3m_{i}m_{j}
(\gamma_{i}^{2}-\gamma _{j}^{2})}\frac{e^{-r/\gamma _{i}}-
e^{-r/\gamma_{j}}}{r}{\bf s}_{i}\cdot {\bf s}_{j}.
\label{hypdd}
\end{eqnarray}

It is possible to treat the electromagnetic potential $U$ as a
perturbation. This procedure is only used for the baryons with the BSS
potential (see Sec.~\ref{sec:bss}).

\section{Results}

\subsection{Determination of the parameters}

We do not want to introduce a lot of new parameters; here we restrict
the number of free parameters to the minimum unavoidable. In particular
the parameters of the strong potential are maintained without
modification. Important information is contained in the electromagnetic
size of the quarks $\gamma _{i}$. This size could depend on the flavor
and on the electrical charge of the quark. If the dependence on charge
were dominant, we could expect that sizes of quarks with the same charge
are similar. Preliminary calculations have shown that the size must be
strongly reduced for heavy quarks, showing that the dependence on mass
must be dominant. Consequently, in order to restrict again the number of
parameters, we assume that the sizes of the $u$ and $d$ quarks are the
same. Thus we impose $\gamma _{u}=\gamma _{d}=\gamma _{n}$. For AL1 and
BSS models, we have $\gamma _{n}$ and $\gamma _{s}$ as free parameters.
An observable that is very sensitive to those parameters is the charge
mean square radius.

In the mesonic sector, the bare charge square radius operator for
pointlike quarks is defined by
\begin{equation}
(r^{2})^{(b)}=\sum_{i=1}^{2}e_{i}\,({\bf r}_{i}-{\bf R})^{2},
\label{r2bop}
\end{equation}
where $e_i$ is the charge for quark $i$, ${\bf r}_{i}$ its position,
and ${\bf R}$ the position of the center of mass. For constituent
quarks, this expression has to be folded with quark density and should
be written instead
\begin{equation}
(r^{2})=\sum_{i=1}^{2}e_{i}\int d{\bf r}^{\prime }\,({\bf r}^{\prime }-
{\bf R})^{2}\,\rho _{i}({\bf r}^{\prime }-{\bf r}_{i}).
\label{r2op}
\end{equation}
Averaging quantity (\ref{r2op}) on the meson wave function provides us
with the charge mean square radius of the meson. Performing the
calculation, one finds that this observable is a sum of a term, that can
be called the bare radius $\left\langle r^{2}\right\rangle ^{(b)}$
(which is essentially the mean value of quantity~(\ref{r2bop}) on meson
wave function), plus a term which is essentially the dipole moment of
the density. With a Yukawa density, one has explicitly
\begin{equation}
\left\langle r^{2}\right\rangle =\left\langle r^{2}\right\rangle
^{(b)}+6\sum_{i=1}^{2}e_{i}\gamma _{i}^{2}.
\label{r2}
\end{equation}
The dynamical contribution to the square radius is entirely contained in
the bare quantity, whose expression is very well known and is not
recalled here.

From relation (\ref{r2}), one sees that the pion radius depends only on
$\gamma_{n}$ and it is used to determine this parameter. The kaon
radius depends on $\gamma _{n}$ and $\gamma _{s}$; since $\gamma _{n}$
has already be determined from the pion, the kaon radius is used to
determine $\gamma _{s}$. In addition, the AL1 potential needs the
further determination of $\gamma _{c}$ and $\gamma _{b}.$ Since the
radii for $D$ and $B$ resonances are not known experimentally, we
determine the corresponding sizes by requiring a smooth behavior versus
the mass. Moreover, the uncertainty on the kaon radius is rather large.
Fortunately, we checked that the isospin splitting does not depend too
much on the precise value of the quark size. Our accepted values for the
electromagnetic sizes are summarized on table~\ref{emsize}.

Other parameters that need to be changed are the quark masses. In fact,
the only important ingredient for the splittings is the mass difference
between the down and up quarks:
$\Delta =\widetilde{m}_{d}-\widetilde{m}_{u}$. In view of this, we
choose to maintain the $s$, $c$, and $b$ quark masses at their non
perturbative value $\widetilde{m}_{i}=m_{i}$, and to keep the average
value of the isospin doublet at its non perturbative value
$(\widetilde{m}_{d}+\widetilde{m}_{u})/2=m_{n}$. The size parameters
being determined once and for all on charge radii, we have only
\emph{one free parameter} $\Delta $ at our disposal to try to reproduce
all the known electromagnetic splittings. To really see that this is a
very big constraint, let us recall that for doing the same job, authors
of Refs.~\cite{godf86,caps87} used 4 free parameters and Genovese
\emph{et al} 2 free parameters.
The $\pi ^{0}$ and $\rho ^{0}$ resonances are considered here as
$n \overline{n}$ systems, composed of fictitious quark and antiquark of
mass $\widetilde{m}_{n}=(\widetilde{m}_{u}+\widetilde{m}_{d})/2 = m_n$.
Doing this, we neglect mixing of $I=0$ and $I=1$ components; this is a
good approximation up to second order in $\Delta/m_n$.
One can imagine several
strategies to determine the parameter $\Delta$. We first remarked that
if we fit
$\Delta $ on the mesons, the baryons were very badly reproduced, while
if we fit it on the baryons, the mesons were spoiled less dramatically.
Moreover, among the baryons, some splittings are more affected than
others by a small change of $\Delta$. Thus, we decided to fit this
parameter on one of the most sensitive and well known splitting, namely
$\Sigma^{-}-\Sigma ^{0}= 4.807\pm 0.04$~MeV.

One aim of our study is to see the respective influence of each
component of the electromagnetic potential (\ref{potemd}). The coulomb
part seems to us unavoidable, so we will consider four different
approximations in the following:
\begin{itemize}
\item C: Electromagnetic potential restricted to the coulomb term
(\ref{couldi}) or (\ref{couldd}) alone;
\item CC: Electromagnetic potential restricted to coulomb and contact
(\ref{contdi}) or (\ref{contdd}) terms alone;
\item CH: Electromagnetic potential restricted to coulomb and hyperfine
(\ref{hypdi}) or (\ref{hypdd}) terms alone;
\item T: Total electromagnetic potential (\ref{potemd}) taken into
account.
\end{itemize}
Each approximation requires its own $\Delta $ parameter, but the
$\gamma _{i} $ parameters can be maintained to their values of Table
\ref{emsize}. The corresponding results are gathered in Table
\ref{delta}. It is worth noting that the study of any of these
approximations is a complete calculation by its own and needs a
computational effort identical to the total treatment.

It is funny that for both potentials, increasing values of $\Delta $ are
obtained with approximations CH, C, T, CC respectively. Is it a property
independent of the strong potential (realistic enough to reproduce
baryon and meson spectra)? We have no answer.

\subsection{Experimental data}

A number of experimental data, concerning the electromagnetic
splittings, exist for both the mesonic and baryonic sectors, and for
both light quarks ($u$, $d$, and $s$) systems and systems containing
at least a heavy quark ($c$ and $b$). The use of AL1 potential allows to
study all data, while BSS is restricted to the light quark domain for
the reasons mentioned in Sec.~\ref{sec:bss}. The
values for the splittings are of order of some MeV and some of them are
known with good accuracy.

From the splitting in the nucleon case and many others, the $d$ quark
mass is presumably larger than the $u$ quark mass. From a naive argument
based on rest masses only, the hierarchy of the splittings can be
explained although the quantitative value needs much refined
explanation. However, contrary to this naive argument, a number of
puzzling questions still arise and some of them are not solved in a
satisfactory manner by the up to date theoretical studies. Let us list
some of them:
\begin{itemize}
\item $n-p=1.293$~MeV is much weaker than
$\pi ^{+}-\pi ^{0}=4.594$~MeV, despite the fact that a very naive quark
model gives $n(udd)-p(duu)=\Delta$ and
$\pi^{+}(u\bar d)-\pi ^{0}([u\bar u - d\bar d]/\sqrt{2})=0$;
\item $\pi ^{+}-\pi ^{0}$ has a large positive value while
$\rho ^{+}-\rho ^{0}$ has a small negative value (a positive value is
compatible with the error bar but with a very low value in any case);
\item $D_2^+-D_2^0 \approx 0$ (or even may be negative) while the values
for other $c \bar n$ states are largely positive;
\item $\Sigma _{c}^{++}$ is the highest level of the multiplet,
in contradiction to what one expects from naive arguments where it
should be the lowest one;
\item All the members of the charmed $\Xi$ verify $\Xi_c^0 > \Xi_c^+$
except the particular states $\Xi_c^{0*} < \Xi_c^{+*}$ which satisfy
the opposite relation.
\end{itemize}
These remarks illustrate the fact that the mass difference between down
and up quarks cannot be the only -- or even leading -- ingredient to
explain the splitting; the internal dynamics also plays an important
role. An important aim of this paper is devoted to examine this
question.

\subsection{Influence of various approximations}

In this part we want to study the effect of using various
approximations C, CC, CH and T of the electromagnetic hamiltonian. An
exact treatment is performed, except for the baryons with BSS potential,
as explained previously.

Let us first present the results obtained with AL1 potential. They are
gathered on Table~\ref{mesal1} for the mesons and on Table~\ref{baral1}
for the baryons. Few comments are in order:
\begin{itemize}
\item  The various approximations differ significantly one from the
other. This proves again that the electromagnetic splitting is an
observable very sensitive to the physical content put in it. A more
quantitative comparison is relegated later on.
\item  Although the experimental data cannot be reproduced with a good
precision, the calculated values have the good order of magnitude and
respect more or less the hierarchy. We mean by this that the order of a
given multiplet is generally the good one and that a large (small)
theoretical splitting corresponds to a large (small) experimental one.
Let us recall that we have only one free parameter $\Delta $, which has
been fitted on the $\Sigma^{-}-\Sigma^{0}$ value.
\item  From time to time the sign is wrong (order is opposite to the
experimental one),
but this effect occurs generally when compatibility is not excluded due
to error bars, or at least when the experimental uncertainty is large.
\item  Let us comment on the puzzling questions raised in the
introduction of this section. The splitting among the pions is larger
than the splitting among the nucleons, but not enough to claim that the
problem is solved. The $\rho^+-\rho^0$ mass difference is still wrong
but the value is lower than the naive expected one. The characteristic
pattern for the $D$ mesons is good but the quantitative values fail. The
order in the $\Sigma _{c}$ multiplet is a good one and the quantitative
value are also very good; this case is really a success (let us mention
that in Ref.~\cite{caps87} the theoretical value agrees well with the
experimental data, but since then the data changed and the agreement is
less good with the new value). The problem for
the $\Xi_c$ is not solved but the order is the good one for 3 resonances
over 4.
\end{itemize}

Let us now have a look on the situation concerning the wave functions
arising from the BSS potential. The mesonic sector is presented in Table
\ref{mesbs2} and the baryonic sector in Table~\ref{barbs2}. The same
comments as for AL1 can be made for BSS generally, but in this case the
puzzle concerning the $\rho $ has found a solution. Unfortunately, the
nucleon splitting is much too large.

Moreover the BSS results are quite different from AL1 results,
indicating that, in order to describe the electromagnetic splitting, not
only a good form for the electromagnetic part of the interaction is
important, but also the strong one via the wave function. This
conclusion is very important.
In order to draw this conclusion, it was necessary to use very different
strong potentials that lead equally good results on spectra both in the
mesonic and in the baryonic sectors. These requirements are not easy to
be met. AL1 and BSS allow such a fruitful comparison.

In order to grasp more quantitatively the influence of each
approximation, as well as to compare the effect of the strong potential,
we calculate a chi-square value on the experimental sample (in fact a
chi-square divided by the number of data in the set). In order to avoid
a too much important weight on very precise values, a minimum
uncertainty at 0.1~MeV has been assigned arbitrarily to those values.
The results are gathered in Table~\ref{chi2}. Let us note that this
chi-square has no precise statistical relevance; it is just a convenient
way to get a synthetic view of the comparison between the various
approximations. To have a more refined
analysis, we separate the sample for meson (denoted by ``M" in the
Table), for baryon (``B" in the Table), or the entire sample of hadrons
(``H" = ``M"+``B" in the Table). Moreover we also distinguish between
light sector ($u$, $d$, $s$ denoted by ``l" in the Table), heavy sector
($c$, $b$ denoted by ``h" in the Table), or all sectors (denoted by ``a"
= ``l"+``h" in the Table). For instance the line ``hB" means a
chi-square calculated on heavy baryons, ``lM" calculated on light
mesons, ``aH" on the entire sample, \ldots

Interesting remarks can be emphasized:
\begin{itemize}
\item  The baryonic sector is explained in a much more satisfactory way,
whatever the chosen strong potential. This is the consequence of our
arbitrary choice to adjust the free parameter $\Delta$ on a baryon
resonance. Should we have chosen to fit $\Delta$ on a meson resonance,
the mesonic sector would have been much best reproduced, but at the
price of a dramatic spoiling of the baryonic sector, and with an overall
worse description.
It appears that a consistent description of electromagnetic splittings
for both mesons and baryons is hardly feasible with a model containing
only one free parameter. Let us stress that our model probably do not
contain all possible sources of splitting. For instance, the pion mass
difference can be explained with the vector meson dominance assumption
\cite{saku69}. Nevertheless, this last model relies on pointlike pion
while mesons are composite particles in our approach. So it is difficult
to compare such so different phenomenological models of QCD.
\item  Concerning the BSS calculation, the exact formalism (T
approximation) is the best one for mesons, but it is the CH
approximation that is the best for baryons, and the C approximation
(presumably the crudest one) for the entire set. Anyhow, addition of the
contact term deteriorates seriously the results.
\item  Concerning the AL1 calculation, several conclusions are drawn.
For baryons, all approximations are roughly of the same quality. What is
gained on data by one approximation versus the others is lost on
another data. But curiously CC is the best approximations and T the
worse one, so that the conclusion is rather different from BSS case. For
mesons, CH is always the best approximation and CC
the worse. For the totality of the sample this conclusion remains,
while the exact treatment (T) is just a bit better than the crudest one
(C). Here again the contact term has a very bad influence.
\item  AL1 and BSS calculations give results of comparable quality for
the whole set, but BSS is much better for mesons, and AL1 much better
for baryons. This may be explained by the fact that only a perturbative
treatment (and not an exact one as in AL1) can be performed with BSS in
the baryonic sector.
\item It is interesting to compare our results with some previous
studies
treating both meson and baryon electromagnetic splittings:
\begin{itemize}
\item In Refs.~\cite{godf86,caps87}, the results look at least as good
as ours, but we
want address two comments. First, some data have changed since that time
(for instance, the old value of 1.8 MeV for the
$\Sigma^{++}_c - \Sigma^0_c$ splitting is now 0.35 MeV) and new ones are
now available (for instance, their calculated value of 4.4 MeV for the
$D^+_2 - D^0_2$ is now experimentally estimated at 0.1 MeV). Second, the
value of the electromagnetic quark size is determined by a best fit on
the splittings, while in our case, the same parameter is fixed by meson
form factors and cannot be considered as a free parameter.
\item In Ref.~\cite{geno98}, the results look also rather good, but
there exist several differences with the present work. First, the
contact term is absent. Second, only the hyperfine term is dressed in
order to avoid a collapse. Third, two free parameters were used, the $u$
and $d$ masses separately.
\end{itemize}
\end{itemize}
No doubt that if we have allowed more free parameters in our model (for
example releasing the constraint
$(\widetilde{m}_{u}+\widetilde{m}_{d})/2 = m_n$), our results would have
appeared better. But this was not the ultimate goal of our paper; we
were interested in discovering what are the necessary ingredients to
explain the splittings.

Another important point for the consistency of our approach is that,
with the new values of the $u$ and $d$ masses, the absolute masses for
the meson and baryon resonances are also well reproduced. This point is
not often studied in previous works. Just to convince the reader that
our formalism is able to provide a good description of meson and baryons
simultaneously, we present below the absolute masses of one member of
the multiplet (the other can be obtained using the values of the
splittings given above) for both mesons (Table \ref{mesabs}) and baryons
(Table~\ref{barabs}). The T approximation is used and the treatment is
exact except for baryons with the BSS potential.

The small discrepancy for light baryons in the case of AL1 potential can
be attributed to a three-body force \cite{bhad81}, which is mass
dependent. The instanton does not give rise to a three-body force for
the baryon \cite{blas90,munz94}, and the parameters have been fitted
directly to the absolute masses of mesons and baryons. The agreement
with experimental data is rather satisfactory.

\section{Conclusions}

In this paper, we calculated the electromagnetic splitting on hadronic
systems. This observable is a very sensitive test of the formalism
because it results from a very subtle and fine balance between several
physical ingredients. In order to concentrate on the physical aspects of
the problem, we were very cautious in the numerical treatment, both for
the two-body and three-body problems. Thus we are very confident with
our numerical results, and interesting conclusions can be emphasized.

As compared to previous works we considered the total electromagnetic
hamiltonian (excepted Darwin, spin-orbit, and tensor forces that we
believe
to play a very minor role). In particular we took into account the
so-called contact term that is usually neglected. Moreover, we treated
the quarks as constituent particles with an electromagnetic size that
modifies the form of the electromagnetic interaction. Lastly, we based
our calculations on wave functions resulting from two different strong
potentials: the phenomenological AL1 potential to be used with a non
relativistic kinematics energy operator and the more fundamental BSS
potential, including instanton effects, to be used with a relativistic
kinetic energy operator. The size of the quarks were determined in order
to reproduce the pion and kaon charge form factors. Only one free
parameter, the mass difference between down and up quarks, is left at
our disposal; it was fitted on the $\Sigma ^{-}-\Sigma ^{0}$ splitting.
We want to emphasize that in this paper, with only one parameter, we try
to reproduce the totality of the experimental data, including mesons and
baryons. To our opinion, this is a condition to pretend to some
consistency in the formalism. This condition is rarely met in previous
calculations, since authors very often restricted themselves either to
mesons (or even to less restrictive domains) or to baryons.

We first showed that dressing the strong and electromagnetic
interaction is a necessity to obtain in a consistent way both the hadron
spectra, the meson form factors and the electromagnetic splitting.

By comparison of AL1 and BSS results, we stressed that the strong
potential, via the wave function, is an important ingredient in the
description of electromagnetic splittings. In our particular case, AL1
gives a better description of baryons, and BSS a better description of
mesons.

But we proved also that the electromagnetic hamiltonian is equally
important for explaining the splittings. Adding or removing a term
(contact or hyperfine) has a non negligible influence. In particular
taking into account the contact term spoils a lot the results, and
curiously it is the approximation based on coulomb+hyperfine (the usual
ingredient for many people) which is globally the best.

In our formalism the splittings are described in a reasonable way,
specially owing to the fact that we have only one free parameter to try
to reproduce 26 experimental data; the
order among multiplet masses is generally correct and the values have
the right order of magnitude. But the agreement is far from being
perfect; some suggested puzzles have been solved but some others are
still opened questions. New improvements must be done in future studies.

\acknowledgments

B. Silvestre-Brac and C. Semay greatly acknowledge the financial support
provided by cooperation agreement CNRS/CGRI-FNRS (France-Belgium). C.
Semay (FNRS Research Associate position) and F. Brau (FNRS Postdoctoral
Researcher position) would like to thank FNRS for
financial support .

\clearpage

\begin{table}[tbp] \centering
\begin{tabular}{|l|c|c|c|c|}
\hline
 & $\gamma _{n}$ & $\gamma _{s}$ & $\gamma _{c}$ & $\gamma _{b}$ \\
\hline
AL1 & $1.225$ & $0.200 $ & $0.040$ & $0.013$ \\
BSS & $1.330$ & $0.450$ & $-$ & $-$ \\
\hline
\end{tabular}
\caption{Electromagnetic sizes $\gamma_f$ of the constituents quarks (in
GeV$^{-1}$), for different flavors $f$, and for the two different
strong potentials. \label{emsize}}
\end{table}

\begin{table}[tbp] \centering
\begin{tabular}{|l|r|r|r|r|}
\hline
    & \multicolumn{1}{|c|}{C} & \multicolumn{1}{|c|}{CC} &
    \multicolumn{1}{|c|}{CH} & \multicolumn{1}{|c|}{T} \\
\hline
AL1 & $13.2$ & $14.4$ & $12.6$ & $13.6$ \\
BSS & $7.2 $ & $9.4$ & $6.2$ & $8.4$ \\
\hline
\end{tabular}
\caption{Values of the parameter
$\Delta =\widetilde{m}_{d}-\widetilde{m}_{u}$ (in MeV), for the two
potentials, and for the four different approximations presented in the
text. \label{delta}}
\end{table}

\begin{table}[tbp] \centering
\begin{tabular}{|l|c|r|r|r|r|}
\hline
\multicolumn{1}{|c|}{Splitting} & Exp & \multicolumn{1}{|c|}{C} &
\multicolumn{1}{|c|}{CC} & \multicolumn{1}{|c|}{CH} &
\multicolumn{1}{|c|}{T} \\
\hline
$\pi ^{+}-\pi ^{0}$ & $4.594\pm 0.001$ & $1.239$ & $0.828$ & $2.111$ &
$1.693$ \\
$\rho ^{+}-\rho ^{0}$ & $-0.5\pm 0.7$ & $1.009$ & $0.857$ & $0.870$ &
$0.713$ \\
$K^{0}-K^{+}$ & $3.995\pm 0.034$ & $8.096$ & $9.290$ & $7.054$ & $8.109$
\\
$K^{0\ast }-K^{+\ast }$ & $6.7\pm 1.2$ & $1.132$ & $1.458$ & $1.139$ & $
1.436 $ \\
$K_{2}^{0}-K_{2}^{+}$ & $6.8\pm 2.8$ & $-0.794$ & $-0.793$ & $-0.763$ &
$-0.757$ \\
$D^{+}-D^{0}$ & $4.78\pm 0.10$ & $2.478$ & $1.977$ & $2.821$ & $2.308$
\\
$D^{+\ast }-D^{0\ast }$ & $2.6\pm 1.8$ & $1.513$ & $1.120$ & $1.428$ &
$1.037 $ \\
$D_{1}^{+}-D_{1}^{0}$ & $6.8\pm 5$ & $-2.038$ & $-2.409$ & $-1.839$ &
$-2.161$ \\
$D_{2}^{+}-D_{2}^{0}$ & $0.1\pm 4$ & $-2.058$ & $-2.411$ & $-1.937$ &
$-2.235$ \\
$B^{0}-B^{+}$ & $0.33\pm 0.28$ & $-1.877$ & $-1.715$ & $-1.891$ &
$-1.713$ \\
\hline
\end{tabular}
\caption{Meson electromagnetic splittings (in MeV) calculated for 4
different approximations C, CC, CH, and T, as explained in the text. The
wave functions result from the strong potential AL1 and the numerical
treatment is exact. The experimental values (Exp) come from
Ref.~\cite{pdg}. \label{mesal1}}
\end{table}

\begin{table}[tbp] \centering
\begin{tabular}{|l|c|r|r|r|r|}
\hline
\multicolumn{1}{|c|}{Splitting} & Exp & \multicolumn{1}{|c|}{C} &
\multicolumn{1}{|c|}{CC} & \multicolumn{1}{|c|}{CH} &
\multicolumn{1}{|c|}{T} \\ \hline
$n-p$ & $1.293$ & $0.89$ & $1.15$ & $0.91$ & $1.15$ \\
$\Delta ^{0}-\Delta ^{++}$ & $2.25\pm 0.68$ & $2.68$ & $3.68$ & $2.79$
& $3.72$ \\
$\Delta ^{+}-\Delta ^{++}$ & $1.2\pm 0.6$ & $0.57$ & $1.22$ & $0.73$
& $1.35$ \\
$\Sigma ^{-}-\Sigma ^{0}$ & $4.807\pm 0.04$ & $4.82$ & $4.82$ &
$4.81$ & $4.76$ \\
$\Sigma ^{-}-\Sigma ^{+}$ & $8.08\pm 0.08$ & $7.87$ & $8.41$ & $8.27$
& $8.55$ \\
$\Sigma ^{-\ast }-\Sigma ^{0\ast }$ & $2.0\pm 2.4$ & $3.26$ & $3.23$ &
$3.04$ & $2.94$ \\
$\Sigma ^{-\ast }-\Sigma ^{+\ast }$ & $0\pm 4$ & $1.71$ & $1.99$ &
$1.68$ & $1.96$ \\
$\Xi ^{-}-\Xi ^{0}$ & $6.48\pm 0.24$ & $7.12$ & $7.21$ & $7.38$ &
$7.38$ \\
$\Xi ^{-\ast }-\Xi ^{0\ast }$ & $3.2\pm 0.6$ & $3.01$ & $2.91$ &
$2.80$ & $2.66$ \\
$\Sigma _{c}^{++}-\Sigma _{c}^{0}$ & $0.35 \pm 0.18$ & $1.00$ & $-0.02$
& $1.35$ & $0.37$ \\
$\Sigma _{c}^{0}-\Sigma _{c}^{+}$ & $0.9\pm 0.4$ & $0.32$ & $0.64$ &
$0.02$ & $0.33$ \\
$\Sigma _{c}^{++\ast }-\Sigma _{c}^{0\ast }$ & $1.9\pm 1.7$ & $1.37$ &
$0.27$ & $1.33$ & $0.19$ \\
$\Xi _{c}^{0}-\Xi _{c}^{+}$ & $5.5\pm 1.8$ & $2.81$ & $3.28$ & $3.01$
& $3.42$ \\
$\Xi _{c}^{0\prime }-\Xi _{c}^{+^{\prime }}$ & $\simeq 4.2\pm 3.5$ &
$0.20$ & $0.49$ & $-0.08$ & $0.20$ \\
$\Xi _{c}^{+\ast }-\Xi _{c}^{0\ast }$ & $\simeq 2.9\pm 2.0$ & $-0.08$ &
$-0.31$ & $-0.03$ & $-0.25$ \\
$\Xi _{c}^{0\ast \ast }-\Xi _{c}^{+\ast \ast }$ & $\simeq 4.1\pm 2.5$ &
$3.09$ & $3.42$ & $3.24$ & $3.51$ \\
\hline
\end{tabular}
\caption{Same as Table~\ref{mesal1} for baryons. The theoretical
uncertainty may affect only the last digit. \label{baral1}}
\end{table}

\begin{table}[tbp] \centering
\begin{tabular}{|l|c|r|r|r|r|}
\hline
\multicolumn{1}{|c|}{Splitting} & Exp & \multicolumn{1}{|c|}{C} &
\multicolumn{1}{|c|}{CC} & \multicolumn{1}{|c|}{CH} &
\multicolumn{1}{|c|}{T} \\
\hline
$\pi ^{+}-\pi ^{0}$ & $4.594\pm 0.001$ & $1.22$ & $-0.21$ & $4.11$ &
$2.97$ \\
$\rho ^{+}-\rho ^{0}$ & $-0.5\pm 0.7$ & $1.00$ & $0.40$ & $0.59$ &
$-0.01 $ \\
$K^{0}-K^{+}$ & $3.995\pm 0.034$ & $1.54$ & $3.68$ & $-0.97$ &
$1.17$ \\
$K^{0\ast }-K^{+\ast }$ & $6.7\pm 1.2$ & $2.88$ & $4.50$ & $2.63$ & $
4.25 $ \\
$K_{2}^{0}-K_{2}^{+}$ & $6.8\pm 2.8$ & $2.41$ & $3.49$ & $2.07$ &
$2.83$ \\
\hline
\end{tabular}
\caption{Meson electromagnetic splittings (in MeV) calculated for 4
different approximations C, CC, CH, and T, as explained in the text. The
wave functions result from the strong potential BSS and the numerical
treatment is exact. The experimental values (Exp) come from
Ref.~\cite{pdg}. \label{mesbs2}}
\end{table}

\begin{table}[tbp] \centering
\begin{tabular}{|l|c|r|r|r|r|}
\hline
\multicolumn{1}{|c|}{Splitting} & Exp & \multicolumn{1}{|c|}{C} &
\multicolumn{1}{|c|}{CC} & \multicolumn{1}{|c|}{CH} &
\multicolumn{1}{|c|}{T} \\
\hline
$n-p$ & $1.293$ & $2.78$ & $4.28$ & $2.62$ & $4.08$ \\
$\Delta ^{0}-\Delta ^{++}$ & $2.25\pm 0.68$ & $4.24$ & $8.21$ & $4.37$
& $8.34$ \\
$\Delta ^{+}-\Delta ^{++}$ & $1.2\pm 0.6$ & $1.18$ & $3.69$ & $1.59$
& $4.09$ \\
$\Sigma ^{-}-\Sigma ^{0}$ & $4.807\pm 0.035$ & $4.83$ & $4.81$ &
$4.81$ & $4.80$ \\
$\Sigma ^{-}-\Sigma ^{+}$ & $8.08\pm 0.08$ & $7.68$ & $8.90$ & $8.47$
& $9.69$ \\
$\Sigma ^{-\ast }-\Sigma ^{0\ast }$ & $2.0\pm 2.4$ & $4.98$ & $5.37$ &
$4.06$ & $4.43$ \\
$\Sigma ^{-\ast }-\Sigma ^{+\ast }$ & $0\pm 4$ & $3.12$ & $4.54$ &
$2.88$ & $4.33$ \\
$\Xi ^{-}-\Xi ^{0}$ & $6.48\pm 0.24$ & $4.82$ & $4.50$ & $5.87$ &
$5.55$ \\
$\Xi ^{-\ast }-\Xi ^{0\ast }$ & $3.2\pm 0.6$ & $5.09$ & $5.38$ &
$4.17$ & $4.46$ \\
\hline
\end{tabular}
\caption{Same as Table~\ref{mesbs2} for baryons. The theoretical
uncertainty may affect only the last digit. Let us recall that, for
technical reasons, the theoretical values were obtained with a
perturbative treatment. \label{barbs2}}
\end{table}

\begin{table}[tbp] \centering
\begin{tabular}{|c|r|r|r|r|r|r|r|r|}
\cline{2-9}
\multicolumn{1}{c|}{\ } & \multicolumn{4}{|c|}{AL1} &
\multicolumn{4}{|c|}{BSS} \\
\hline
$\chi^2$ & \multicolumn{1}{|c|}{C} & \multicolumn{1}{|c|}{CC} &
\multicolumn{1}{|c|}{CH} & \multicolumn{1}{|c|}{T} &
\multicolumn{1}{|c|}{C} & \multicolumn{1}{|c|}{CC} &
\multicolumn{1}{|c|}{CH} & \multicolumn{1}{|c|}{T} \\
\hline
lM & $568$ & $851$ & $317$ & $513$ & $350$ & $464$ & $500$ & $213$ \\
lB & $3.3$ & $3.0$ & $3.8$ & $5.0$ & $34$ & $127$ & $23$ & $129$ \\
lH & $205$ & $306$ & $116$ & $186$ & $147$ & $247$ & $194$ & $159$ \\
\hline
hM & $119$ & $169$ & $90$ & $134$ & $-$ & $-$ & $-$ & $-$ \\
hB & $3.0$ & $1.5$ & $5.9$ & $1.2$ & $-$ & $-$ & $-$ & $-$ \\
hH & $51$ & $73$ & $41$ & $56$ & $-$ & $-$ & $-$ & $-$ \\
\hline
aM & $344$ & $510$ & $204$ & $324$ & $-$ & $-$ & $-$ & $-$ \\ \
aB & $3.2$ & $2.4$ & $4.7$ & $3.3$ & $-$ & $-$ & $-$ & $-$ \\
aH & $134$ & $198$ & $81$ & $127$ & $-$ & $-$ & $-$ & $-$ \\
\hline
\end{tabular}
\caption{Chi-square values divided by the number of data in the sample.
The meaning of approximations C, CC, CH, and T have been explained in
the text. The meaning of the rows is related to the sub-samples taken
into account: ``l", ``h", ``a" for light, heavy and all sectors
respectively; ``M", ``B", ``H" for meson, baryon, hadron sectors
respectively. \label{chi2}}
\end{table}

\begin{table}[tbp] \centering
\begin{tabular}{|c|r|r|r|}
\hline
System & \multicolumn{1}{c|}{Exp} & \multicolumn{1}{c|}{AL1} &
\multicolumn{1}{c|}{BSS} \\
\hline
$\pi ^{0}$ & $134.98$ & $137.3$ & $145.8$ \\
$\rho ^{0}$ & $769.0$ & $769.7$ & $756.2$ \\
$K^{+}$ & $493.68$ & $486.6$ & $491.2$ \\
$K^{+\ast }$ & $891.66$ & $902.7$ & $888.7$ \\
$K_{2}^{+}$ & $1425.6$ & $1333.4$ & $1377.5$ \\
$D^{+}$ & $1869.3$ & $1863.2$ & $-$ \\
$D^{+\ast }$ & $2010.0$ & $2016.4$ & $-$ \\
$D_{1}^{+}$ & $2427$ & $2419.6$ & $-$ \\
$D_{2}^{+}$ & $2459$ & $2451$ & $-$ \\
$B^{+}$ & $5279$ & $5295$ & $-$ \\ \hline
\end{tabular}
\caption{Absolute masses for some mesons. Calculations are
done with strong potentials AL1 and BSS, and with the total
electromagnetic hamiltonian. The experimental values (Exp) come from
Ref.~\cite{pdg}. \label{mesabs}}
\end{table}

\begin{table}[tbp] \centering
\begin{tabular}{|c|r|r|r|}
\hline
System & \multicolumn{1}{c|}{Exp} & \multicolumn{1}{c|}{AL1} &
\multicolumn{1}{c|}{BSS} \\ \hline
$p$ & $938$ & $994$ & $935$ \\
$\Delta ^{0}$ & $1234$ & $1308$ & $1260$ \\
$\Lambda $ & $1116$ & $1149$ & $1105$ \\
$\Sigma ^{-}$ & $1197$ & $1233$ & $1201$ \\
$\Sigma ^{-\ast }$ & $1387$ & $1439$ & $1395$ \\
$\Xi ^{-}$ & $1321$ & $1343$ & $1323$ \\
$\Xi ^{-\ast }$ & $1535$ & $1560$ & $1522$ \\
$\Omega $ & $1672$ & $1675$ & $1646$ \\
$\Lambda _{c}^{+}$ & $2285$ & $2290$ & $-$ \\
$\Sigma _{c}^{++}$ & $2453$ & $2466$ & $-$ \\
$\Xi _{c}^{+}$ & $2466$ & $2467$ & $-$ \\
$\Xi _{c}^{+\prime }$ & $2574$ & $2572$ & $-$ \\
$\Xi _{c}^{+\ast }$ & $2647$ & $2650$ & $-$ \\
$\Xi _{c}^{+\ast \ast }$ & $2815$ & $2788$ & $-$ \\
$\Omega _{c}^{0}$ & $2697$ & $2675$ & $-$ \\
$\Lambda _{b}^{0}$ & $5624$ & $5635$ & $-$ \\
\hline
\end{tabular}
\caption{Same as Table~\ref{mesabs} for baryons. \label{barabs}}
\end{table}

\end{document}